\documentclass[11pt,preprint]{aastex}
\usepackage{times}
\usepackage{graphicx}

\let\oldfootsep=\footnotesep
\newcommand\ltsima{$\; \buildrel <\over\sim \;$}
\newcommand\simlt{\lower.5ex\hbox{\ltsima}}
\newcommand\gtsima{$\; \buildrel >\over\sim \;$}
\newcommand\simgt{\lower.5ex\hbox{\gtsima}}
\newcommand\etal{et~al.}

\def\ten#1{\times 10^{#1}}
\newcommand\msun { \rm {M_\odot}}
\newcommand\that{{\widehat t}\,} 

\newcommand\eff{{\cal E}}

\newcommand\pac{Paczy{\'n}ski }
\newcommand\ie{{\it i.e. }}

\newcommand\kpc {\, {\rm kpc}}
\newcommand\Nexp{N_{\rm exp}}
\def\VEV#1{\left\langle #1\right\rangle}

\newcommand\prii{^{\prime\prime }}

\shorttitle{}
\shortauthors{Bennett}

\begin{document}


\title{Large Magellanic Cloud Microlensing Optical Depth with
         Imperfect Event Selection}


\author{David~P.~Bennett  } 
\affil{Department of Physics,
    University of Notre Dame, IN 46556, USA}
\email {bennett@nd.edu}


\clearpage

\begin{abstract}
I present a new analysis of the MACHO Project 5.7 year Large 
Magellanic Cloud (LMC) microlensing data set that incorporates
the effects of contamination of the microlensing event sample
by variable stars. Photometric monitoring of MACHO LMC microlensing
event candidates by the EROS and OGLE groups
has revealed that one of these events is likely to
be a variable star, while additional data has confirmed that many of the
other events are very likely to be microlensing. This additional data on
the nature of the MACHO microlensing candidates is incorporated into
a simple likelihood analysis to derive a probability distribution for the 
number of MACHO microlens candidates that are true microlensing
events. This analysis shows that 10-12 of the 13 events that passed the
MACHO selection criteria are likely to be microlensing events, with the
other 1-3 being variable stars. This likelihood analysis is also
used to show that the main conclusions of the MACHO LMC analysis
are unchanged by the variable star contamination. The microlensing
optical depth toward the LMC is $\tau  = 1.0\pm 0.3 \times 10^{-7}$. If
this is due to microlensing by known stellar populations, plus an additional
population of lens objects in the Galactic halo, then the new halo population
would account for 16\% of the mass of a standard Galactic halo.
The MACHO detection
exceeds the expected background of 2 events expected from ordinary
stars in standard models of the Milky Way and LMC at the 99.98\% confidence
level. The background prediction is increased to 3 events if maximal disk
models are assumed for both the Milky Way and LMC, but this model
fails to account for the full signal seen by MACHO at the 99.8\% confidence
level.
\end{abstract}




\section{Introduction}
\label{intro}
An important difficulty inherent to gravitational microlensing surveys
is the problem of selecting microlensing events against the background
of variable stars. The chief difficulty with microlensing event selection
is that our a priori knowledge of the background of other types of
variable stars that might mimic microlensing events has been
poorly understood. Fortunately, gravitational microlensing light
curves are usually described by a very simple model 
\citep{pac86} to very high accuracy, and the peak magnifications
is such that some microlensing events are observed with a high
enough signal-to-noise ratio (S/N) to be reasonably convincing
\citep{macho-nat93}, even without a good understanding of the
variable star background. However, many of the conclusions 
from microlensing event samples are statistical in nature, so it
is advantageous to identify as many microlensing events as
possible to improve the statistics. This is particularly true for
microlensing searches observing the Magellanic clouds where the
microlensing optical depth is more than an order of magnitude 
smaller than it is towards the Galactic bulge. 

This need for higher statistics has led to the announcement of
microlensing candidates that have since been rejected. Two of the
first three MACHO LMC microlensing candidates \citep{macho-lmc1}
are no longer considered to be viable microlensing candidates.
MACHO-LMC-2 has exhibited additional variations that are
appear inconsistent with any reasonable microlensing interpretation,
and the MACHO Project has found that candidate MACHO-LMC-3
has a signal-to-noise ratio (S/N) so low that it no longer stands out
from the false microlensing--like signals caused by variable stars
and/or photometry problems \citep{macho-lmc2}. Similarly, the 
first four EROS LMC microlensing candidates 
\citep{eros93,eros-lmc-tau} have also been rejected
\citep{eros2-var,eros12-spec,jetz-mil-tiss,tiss-mil} as likely
variable stars. 

A major advance in the identification of Magellanic Cloud
microlensing events was made by the MACHO Project
in \citet{macho-lmc5.7}
(hereafter A00), where they identified background supernovae
as the dominant non-microlensing background, and developed a
method to remove this background from the microlensing sample.
A similar method has also been implemented by EROS
\citep{jetz-mil-tiss,tiss-mil} to remove supernovae contamination
from their data, and this has led to the rejection of several previous
microlensing candidates.

The question remains, however, whether there is any additional
population of non-microlensing events that still contaminates the
LMC microlensing candidate samples after the supernovae have
been removed. There are some reasons to
believe that this possible contamination is not large. There are now
a number of events that would be quite difficult to explain with another
variability mechanism. These include high magnification events, such as
MACHO-LMC-5 \citep{macho-hstlmc5,dck-lmc5,lmc5-mass}
and MACHO-99-LMC-2
\citep{moa-himag-obs}, caustic crossing binary events
such as MACHO-LMC-9 \citep{macho-lmc9,macho-binaries}, and 
MACHO-98-SMC-1 \citep{joint-98smc1}, and events with
high precision follow-up photometry, such as MACHO-LMC-4, 13, 14
and 15 \citep{macho-96lmc2,lmc-conf}. Also, observations of 
red clump giant stars in both the LMC and Galactic bulge have 
revealed no examples of variability resembling microlensing
\citep{macho-blg45,eros-blg-tau,
pop-blgclump,macho-blg-ev,sumi-ogle-tau}.

Recent observations of microlensing candidate MACHO-LMC-23 by
the EROS \citep{glicens-hawaii,jetz-mil-tiss,tiss-mil} and OGLE
(Udalski, private communication) indicate a subsequent brightening 
of this star approximately 2500 days after the one observed by MACHO.
This could conceivably be explained with a binary source or binary 
lens microlens model, but the MACHO data for the first brightening
episode also does not fit a standard microlensing model very
well \citep{lmc-conf}, and a different modification of standard
microlensing would be required to explain the first brightening
episode. It seems more likely that MACHO-LMC-23 is a variable
star instead of a microlensing event, and I will assume that this is
the case in the rest of this paper.

\citet{massimo-livio} have suggested a number of possible variable
star types that could be mistaken for microlensing events, and it is
possible that one of these types, old novae, could explain both both
MACHO LMC-23 as well as MACHO LMC-2. Another possibility is that
these events are related to the ``blue bumpers" \citep{blue-var} which
have been observed to exhibit multiple brightening episodes for
brighter source stars \citep{macho-lmc1,macho-lmc2,macho-lmc5.7}.
These blue bumpers are thought to be related to Be-stars, and it
seems likely that the EROS-LMC-1 event is an example of this since
the source star is a B-star which exhibits emission lines 
\citep{eros12-spec} and as also exhibited a second brightening 
episode \citep{tiss-mil}. EROS-LMC-2 and 3 have also 
apparently exhibited subsequent variations \citep{eros-lmc-tau,tiss-mil},
but EROS-LMC-2 exhibits significant variability on the light curve
baseline \citep{eros2-var}, and EROS-LMC-3 has significant
light curve variations in the initial light curve peak that may
not be consistent with a microlensing model. So, it is not
clear that that MACHO analysis would have selected
any of the first three EROS events as microlensing candidates.
So, while there are some clues to the physical nature of these 
false microlensing events, it is not clear that they are all caused
by the same mechanism, nor are the details of the candidate
mechanisms well understood. Therefore, my analysis will
be limited to a statistical analysis of the possible level of
contamination of the MACHO sample.

We have seen that
there is strong evidence that many of the MACHO LMC microlensing
candidates are indeed true microlensing events, but one of these events
now appears to be a variable star. In addition, there are several other
of the MACHO LMC microlensing candidates from A00 where the
verdict is uncertain. In Sec.~\ref{sec-eval-mlens}, I present a classification
of the microlensing candidates from A00 that sorts them into three
categories: confirmed, rejected, and unconfirmed. A simple likelihood
method is then used to assign a true microlensing probability to
the unconfirmed events. In Sec.~\ref{sec-mlens-tau-rate}, I discuss how to
use the results of Sec.~\ref{sec-eval-mlens} to produce a new estimate
of the microlensing optical depth, and to compare to models of the
microlensing background due to lensing by ordinary stars in the 
Milky Way and LMC. This section also includes a correction to the
microlensing event detection efficiencies used in A00, where there
was a slight error in the normalization of the LMC luminosity
used in the efficiency calculations. Finally, in Sec.~\ref{sec-conclude},
I discuss the implications of these calculations and the status
of the observed microlensing excess towards the LMC.

\section{Evaluation of Candidate Microlensing Events}
\label{sec-eval-mlens}

A00 presented two sets of candidate microlensing events, which were
selected with selection criteria A and B. Event set A included 13 events
(numbered 1, 4, 5, 6, 7, 8, 13, 14, 15, 18, 21, 23 and 25), and set B,
which was designed to be more inclusive, included all the set A events
plus events 9, 20, 22, and 27. Additional data obtained for event 22 
shortly before the A00 paper went to press indicated the source star
was barely resolved in $0.9\prii$ seeing and that this source had the
spectrum of an active background galaxy. This strongly suggests that 
the event was with a very long duration supernova in the background
galaxy or some type of active galactic nucleus  phenomena. As mentioned
above, event 23 now appears to be a variable star
\citep{glicens-hawaii,jetz-mil-tiss,tiss-mil,lmc-conf}, so 2 of the A00 events
can be considered to be rejected. However, only event 23 was selected
by the more restrictive cut A, which put tighter constraints on the 
quality of the light curve fit to a standard \pac\ light curve shape. The
drawback of criteria A is that it does reject events like the binary
caustic crossing event 9 \citep{macho-lmc9}. The observed light curve
features of this event are unique to binary caustic crossing events, and this
means that this event is also almost certainly a true microlensing event.
However, these same features also mean that this event has a poor
fit to the standard \pac\ light curve model, and this causes this event to
fail the cuts for selection criteria A. I will return to the effect of this 
systematic error in Sec.~\ref{sec-conclude}. Our focus on the events
selected by criteria A also means that we will not consider events
20 and 27 have the lowest S/N of all the A00 microlensing candidates.

\subsection{Confirmed Microlensing Events}
\label{sec-confirm}

My classification of the 13 MACHO LMC microlensing candidates selected
by criteria A of A00 is listed in Table~\ref{tab-that}. There are three 
different categories of confirmed microlensing candidates listed in this
table: follow-up, clump giant, and lens ID. 

Event 5 is the only candidate confirmed with a direct lens identification
\citep{macho-hstlmc5}. The final analysis of this event shows
\citep{dck-lmc5,lmc5-mass} that the microlensing data predicts a lens
mass and distance that are completely consistent with the stellar
brightness, color, and parallax. So, this event has passed three
independent tests, and its microlensing interpretation
is considered to be essentially certain.

Events 4, 13, 14, 15 all have high precision follow-up photometry from 
the CTIO 0.9m telescope obtained in seeing that is typically much
better than in the MACHO survey images
\citep{macho-96lmc2,lmc-conf}. The improved seeing and longer
exposure times for the CTIO data yields photometry that has 
estimated uncertainties 2.5-5 times smaller than the original
MACHO data, and in each case the microlensing interpretation
is confirmed. Events 16 and 17, were also discovered by the
MACHO alert system \citep{macho-alert}
as alerts 97-LMC-4 and 98-LMC-1, 
but these were classified as supernovae by the MACHO 5.7 
year analysis. MACHO Alerts 97-LMC-2, 3 and 5 were
also in the MACHO 5.7 year data set, but these light curves
failed other cuts and made neither the microlensing or
supernovae samples. MACHO Alerts 97-LMC-6 and 7
occurred in fields not included in the 5.7 year analysis.
The only microlensing candidates from the 5.7 year analysis
that occurred when the MACHO alert system was operating are
events 18, 20, 22, 23, and 25. However, events 22, 23 and 25 occurred
at times when the alert system was not running in their
respective fields. Event 18 reached maximum
magnification in the middle of a 40-day period when the
MACHO alert system was not operating due to a camera
upgrade and the subsequent need to modify photometry 
code input files. Event 20 was identified by the MACHO alert
system, but it did not pass the subjective selection criteria
for an alert. However, event 20 did not pass selection criteria A,
and so this is not relevant to our analysis. The only criteria A
events that occurred when the MACHO alert system was running
were the events 4, 13, 14, and 15 which have the follow-up
photometry, which has been used to confirm the microlensing
interpretation of these events. Thus, it is fair to consider these
4 events as a randomly selected sub-sample of the events
passing criteria A.

My final sub-category of ``confirmed" events are those event
with red clump giant source stars with achromatic light curves,
and events 1 and 25 fall in this category. Event 1, of course, was
the first event published by MACHO \citep{macho-nat93}, and its
light curve was observed with a high enough S/N to be convincing
to most astronomers, as well as to the MACHO team itself, which
referred to it as the ``gold plated event\rlap." This event does
show a significant deviation from the standard \pac\ light curve,
which is explained by a binary lens model 
\citep{dominik94,rhie_dm96,macho-binaries}.
Event 25 is less dramatic,
but it is classified as confirmed event because of its position on the
red clump giant region of the color magnitude diagram and the fact
that brightening is achromatic. (The chromatic brightening of a red
clump star could be due to blending with a variable main sequence
star.) High confidence in the microlensing interpretation of the
events with red clump giant source stars is justified MACHO
results for Galactic bulge microlensing, where 100\% of $\sim 40$
red clump giant source stars that triggered the MACHO alert 
system with an apparently achromatic brightening in a single
image were later classified as microlensing events based upon
excellent fits to microlensing models using the MACHO survey
data as well as follow-up observations from GMAN
\citep{macho-95blg30} and MPS
\citep{mps-98smc1}.

The situation is quite different for main sequence source stars.
It has long been noticed that there is a class of main sequence
stars in the LMC, sometimes referred to as ``bumpers",
with variability characteristics that are similar to
microlensing \citep{eros-lowmass, eros-lmc-Be, macho-lmc1}.
That is, they exhibit relatively brief brightening episodes,
but spend most of the time at a constant brightness. 
MACHO and EROS attempted to prevent these variable stars
from contaminating their microlensing samples by applying
cuts to the color-magnitude diagrams of the source stars
\citep{eros-lowmass,eros-lmc-tau,eros-smc-tau,
macho-lmc1,macho-lmc2,macho-lmc5.7}, but the evidence indicates
that variable stars that have contaminated previous LMC microlensing
candidate samples \citep{eros12-spec,eros2-var} have properties
similar to the brighter main sequence stars that are removed from
the microlensing search analysis \citep{eros-lmc-Be,blue-var}.
So, it appears that the contamination of microlensing event samples
by variable stars is only a serious problem for source stars on or near
the main sequence.

\subsection{True Microlensing Event Fraction}
\label{sec-true-mlens}

With the classification of microlensing event candidates listed in 
Table~\ref{tab-that} and described in Sec.~\ref{sec-confirm}, it is
possible to estimate the probability that the events classified as
unconfirmed (events 6, 7, 8, 18, and 21) are actually microlensing
events. Of 13 microlensing candidates, candidates, 4 have been
observed in follow-up mode with much higher precision photometry,
which confirms the microlensing interpretation.
Since the selection of these 4 events was independent any other
factor that might relate to event classification, these 4 events are
a fair sub-sample of the full sample of 13 events. This is not the
case for the two red clump giant source star events, since they are
the only two such events. Event 5 is similar to the 4 events with
follow-up photometry in that it appears to have a normal, main
sequence source star, but it is special because its lens star has
been identified. It may be the case that the lens stars for other
events can be identified, but event 5 certainly stands out because
its lens star is bright and too red for a normal LMC star. Thus, the 
conservative choice is not to include event 5 in the fair sub-sample
of events to compare to the unconfirmed events. 

It is certainly the case that event 23 has been singled out as the
one event for which there is evidence of variation that is likely
not to be due to microlensing, and so, it seems reasonable not to
include it in the fair sub-sample of events that will be used in the
comparison to the unconfirmed events. However, this would
minimize the effect of the rejection of this event in the analysis below,
and so I choose to include the rejected event as part of the
fair sub-sample of events even if it should not strictly be 
allowed.

With the identification of this fair sub-sample of events, it is straight
forward a likelihood function for the true microlens fraction, $p$, of 
the events with main sequence source stars, selected by criteria A:
\begin{equation}
\label{eq-like-mlf}
{\cal L}_{\rm ml} \propto p^4\, (1-p) \ .
\end{equation}
This is simply the relative probability of 4 confirmed and 1 rejected
microlensing events in the fair sub-sample of events. This function
is plotted in Fig.~\ref{fig-mlens_prob}. Also plotted in 
Fig.~\ref{fig-mlens_prob} are the likelihood functions for different
choices of the sub-sample. These are 4/4 events confirmed, which
is appropriate for a sub-sample that doesn't include event 23, and
5/6 events confirmed, which refers to a sub-sample including both
event 5 and event 23.

We can apply this likelihood function to the question of how many
of the MACHO events passing criteria A are actually microlensing
events. The appropriate prior distribution of $p$ for Baysean
likelihood analysis is a uniform distribution, and with that prior the
likelihood function in eq.~\ref{eq-like-mlf} and FIg.~\ref{fig-mlens_prob}
can be interpreted as a probability distribution for the true 
microlensing fraction, $p$. This probability distribution can then
be convolved with the binomial distribution for the 5 unconfirmed
events to yield the microlensing event probabilities listed in 
Table~\ref{tab-true}. This table also gives the probabilities that
would be obtained we had assumed that 4/4 or 5/6 sub-sample
events were confirmed to be microlensing.

\section{Microlensing Optical Depth and Rate}
\label{sec-mlens-tau-rate}

In this section, I will use the likelihood function, eq.~\ref{eq-like-mlf}, to
determine a new microlensing optical depth for the MACHO LMC
data, and to compare the observed LMC microlensing rate to the
rate predicted due to microlensing by normal stars in the Milky
Way and LMC. But, I must first correct a systematic error in the
microlensing detection efficiencies determined by the MACHO
Collaboration.

\subsection{Efficiency Normalization Correction}
\label{sec-overcompl}

One known systematic error that was not removed from the A00 analysis
involved the normalization of the underlying stellar luminosity function,
determined from HST images, to the apparent luminosity function as
observed in the MACHO images. This was done by adjusting the
star density to the assumed true luminosity function so that the number 
of stars in magnitude bin $17.5 < V < 18.5$ will match the stellar density
in the same magnitude range seen by MACHO. This magnitude bin was
selected because it is bright enough so that the MACHO images are
not strongly affected by blending, but there are still enough stars to
for accurate sampling in the HST frames. The accuracy of this
procedure was tested by constructing three synthetic images with
parameters chosen to match the MACHO images \citep{macho-lmc5.7eff}.
We found that the number of ``objects" identified in the simulated 
MACHO images was generally larger than the number of input
stars: \ie  
$N_{\rm objects}(17.5 < V < 18.5) > N_{\rm stars}(17.5 < V < 18.5) $.
The synthetic images had ratios of 
$N_{\rm objects}/N_{\rm stars} = 1.04 \pm 0.08$, $1.04 \pm 0.05$, and 
$1.18 \pm 0.04$,
where the errors are 1$\sigma$ Poisson errors. Thus, the MACHO images
are expected to have a formal completeness $> 1$ in this magnitude range.

This overcompleteness is easily explained in terms of blending: the extra
``objects" seen in the ground-based images are simply blends of 
multiple fainter stars within the ground-based seeing disk. Since the luminosity
function rises at fainter magnitudes, the probability of two fainter stars moving
into the $17.5<V<18.5$ bin is larger than the probability of two $17.5<V<18.5$
stars moving out of this bin due to blending. Since the density of $17.5<V<18.5$
stars is relatively low, a linear correction will suffice to correct this problem. The 
correction will depend on the apparent density of $17.5<V<18.5$ stars, 
$N_{\rm objects}$, and upon the 
seeing, which we denote by the number
density of seeing disks, $N_{\rm sd}$. (The area of a seeing disk is defined as
$\pi ({\rm FWHM})^2/4$.) The expression for the number density of real (\ie resolved
by HST) stars in the $17.5<V<18.5$ magnitude range is
\begin{equation}
\label{eq-overcc}
N_{\rm stars} = N_{\rm objects} \left( 1 + b\, {N_{\rm objects}\over N_{\rm sd} } \right)^{-1} \ ,
\end{equation}
where the coefficient $b = 5.6\pm 1.0$ was determined by a linear fit to 
the synthetic image data. The correction factor can be calculated on
a field-by-field basis because each field has an approximately constant
star density and an approximately constant PSF FWHM over its
template image. The resulting corrections range from about 1.5\% in
the lowest density fields to 4\% for a more typical field, to an extreme
value of 16.4\% in field 78. (It was the parameters from this field that
gave the empirical correction of $18\pm4\,$\% in the synthetic image.)

\begin{figure}
\plotone{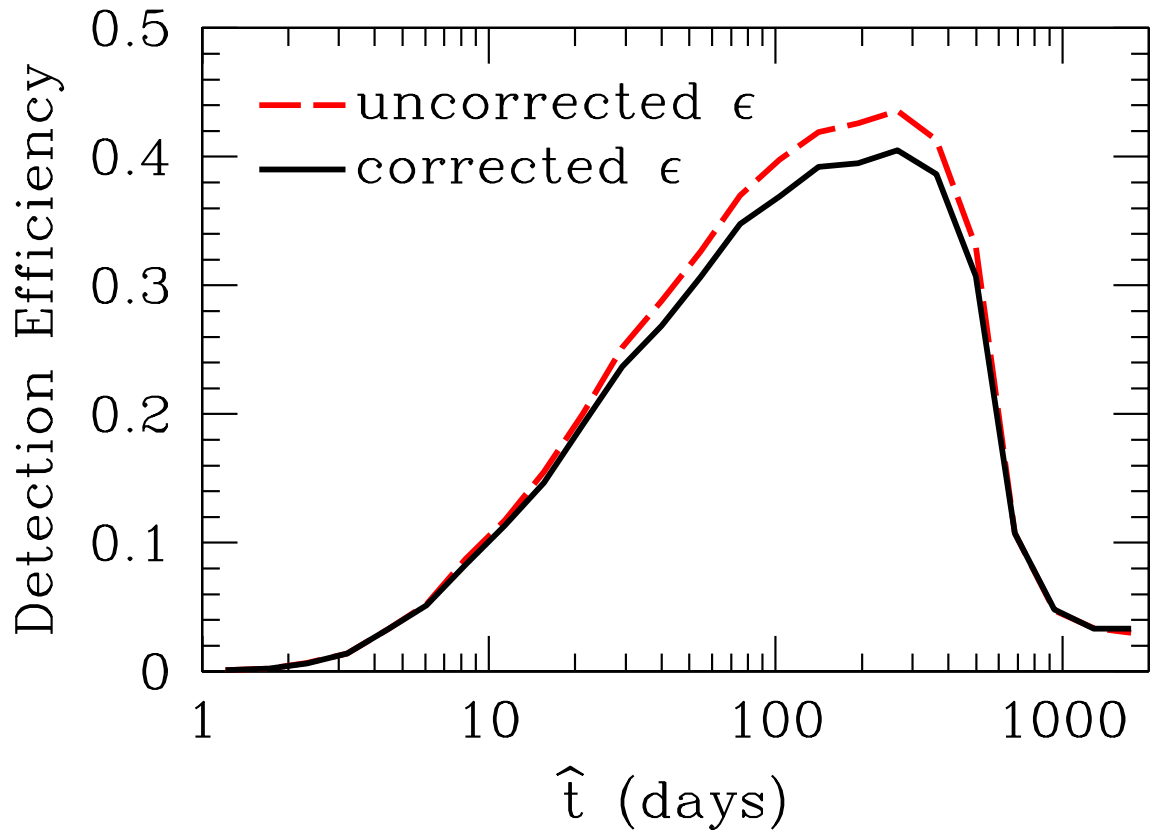}
\caption{
The detection efficiencies for selection criteria A of the MACHO LMC
5.7 year analysis are compared with and without the correction for the
MACHO star list overcompleteness in the $17.5 < V < 18.5$ magnitude
bin.
\label{fig-eff}}
\end{figure}

Although this correction was determined during the preparation of A00,
it was not used in that paper. The reason for this was that there were
thought to be other small systematic errors that might bias the results in
the opposite direction. However, in this paper, I am directly including the
largest of these systematic errors, the uncertainty in the identification of
microlensing events, so it is best to correct for this systematic error,
as well. The overall detection efficiency numbers are corrected by
going back to the field-by-field efficiency values that were determined 
as an intermediate step of the efficiency calculation
\citep{macho-lmc5.7eff}, and then applying the correction formula,
\ref{eq-overcc} to the normalization for each field. Fig.~\ref{fig-eff} 
shows a comparison of the original and corrected efficiencies.
The corrected efficiencies are about 7\% smaller than those reported
in A00 for the timescales of the observed events: 
$30{\,\rm days} < \that < 130{\,\rm days}$. So, this correction, by itself,
would increase the microlensing optical depth measured by MACHO
by about 7\%.

\subsection{ Optical Depth Estimates }
\label{sec-tau}

The microlensing optical depth, $\tau$, is just the fractional area of 
the sky in the direction of our target that is covered by the Einstein 
ring disks of the lens objects. The Einstein ring corresponds to the 
circular image of a point source in perfect alignment behind the 
lens. Conventionally, the Einstein ring is supposed to
lie at the distance of the lens, and the radius is given by
\begin{equation}
\label{eq-re}
R_E = 2\sqrt{G M D_s x(1-x)\over c^2} \ ,
\end{equation}
where $x\equiv D_l/D_s$ and M is the lens mass. $D_l$ and $D_s$ are
the lens and source distances, respectively. The microlensing magnification
is $A > 1.3416$ if there is lens object less than $R_E$ from the line of sight 
to a given source star, so an alternate definition of the microlensing 
optical depth, $\tau$, is the fraction of source stars that are magnified 
by $A > 1.3416$ at any given time. Thus, an experimental estimate of the
microlensing optical depth is given by
\begin{equation}
\label{eq-taumeas}
\tau_{\rm meas} = {1 \over E} {\pi\over 4}
\sum_i {\hat t_i \over \eff(\that_i)} \ ,
\end{equation}
\noindent
where $E=6.12 \ten{7}$ object-years is the total exposure of the MACHO
data set, $\that_i$ is the Einstein ring diameter crossing time of 
the $i$th event, and $\eff({\that_i})$ is its detection efficiency.
As discussed in A00, blending complicates the issue of determining
the correct $\that$ values, and following A00, I use
the statistically corrected values of the blended durations 
$\that_{st}$ (Table~\ref{tab-that}). Note that this choice is
only optimal for optical depth calculations, and it is not
appropriate for studies of the $\that$ distribution itself
because the distribution of $\that_{st}$ is artificially
narrow as noticed by \citet{gj-that-dist} and \citet{rahvar-lmc-stat}.
The $\that$ values determined by fits with blending should be
more accurate on an event-by-event basis, but they can be biased
toward very large $\that$ values due to a light curve degeneracy
involving highly blended events of high magnification and long
duration. The use of the $\that_{st}$ values avoids this bias.
It is also convenient to define the function
\begin{equation}
\label{eq-tau1}
\tau_1(\that_{st}) = {1 \over E}{\pi\over 4}{\that_{st} \over \eff(\that_{st})}\ ,
\end{equation}
which is the contribution to $\tau_{\rm meas}$
from a single observed event with timescale $\that_{st}$. $\tau_1$ values
for each event are listed in Table~\ref{tab-that}.

The final microlensing optical depth depends on the number of 
candidate events that are actually microlensing, and Table~\ref{tab-true}
also shows how the microlensing optical depth changes with the assumed
number of events. Since we don't know which of the unconfirmed events
are the actual microlensing events, we have just assumed that there is
an equal probability for each unconfirmed event to be rejected calculating
the $\tau_N$ values in Table~\ref{tab-true}. The most likely
value for $\tau$ is simply $\tau = 0.99\times 10^{-7}$, which is the
value for the most likely number of true microlensing events, 11.

Table~\ref{tab-true} doesn't reflect the
all the uncertainty in the measured values of $\tau$ because Poisson 
fluctuations in the number detected events have not been included.
As pointed out in \citet{dpb-md_dm} and \citet{hangould-taustat}, 
the standard Poisson statistical formulae cannot be used because 
the different events contribute to $\tau$ with different weights.
Following \citet{dpb-md_dm} and \citet{macho-blg45}, I account for
this with a Monte Carlo method. The number of expected events,
$N_{\rm exp}$, is taken to be a variable, and we construct 
simulated event sets with $\that_{st}$ values drawn from the
observed distribution, and determine what fraction, $f$, of the
simulated data sets has a $\tau$ value that exceeds the measured
value. If $f=0.84$ for a particular assumed $N_{\rm exp}$ value, then
that particular $N_{\rm exp}$ value corresponds to the 1--$\sigma$ 
upper limit on $\tau$, which is listed as the $0.84$ confidence level
in Table~\ref{tab-taucl}. Each $N_{\rm exp}$ can be converted to a
microlensing optical depth by $\tau_{\rm tot} = N_{\rm exp} \, \tau_1$,
where the mean values of $\tau_1$ is assumed.

The procedure described in the previous paragraph was used for the
optical depth estimates in \citet{macho-lmc2} and A00, but it must
be modified here to include the event identification uncertainty. First,
the $\that_{st}$ distribution used for the Monte Carlo weights the
unconfirmed event $\that_{st}$ with 80\% of the weight of the 
confirmed events. Then, the comparison to the observed $\tau$
value uses not the most likely value, $\tau = 0.99\times 10^{-7}$,
but the distribution given in Table~\ref{tab-true}. This method effectively
combines the Poisson uncertainties with the event identification
uncertainty and results in the $\tau$ confidence levels given in
Table~\ref{tab-taucl}. So, the final LMC microlensing optical depth
value is $\tau = (0.99\pm 0.33)\times 10^{-7} $ with 1--$\sigma$ errors 
and $\tau = (0.99 {+ 0.83\atop -0.55})\times 10^{-7}$ with 2--$\sigma$ 
errors. The 95\% confidence level lower limit is 
$\tau > 0.50\times 10^{-7}$.

For comparison, Table~\ref{tab-taucl} also shows the optical depth
confidence intervals for selection criteria A from A00 as well as
the confidence intervals that would result from using only the
4 events with follow-up data to estimate the number of true
microlensing events.

\subsection{Microlensing by Known Stellar Populations}
\label{sec-back-comp}

Table~\ref{tab-stars} shows the expected microlensing properties of
the known stellar populations along the line of site to the LMC as
discussed in A00. The first four lines give the numbers for the
standard models of the Milky Way and LMC, and the fifth line
gives the total microlensing optical depth, rate, and number of
expected events for these standard models. An alternative to the
standard Milky Way disk model is the maximum disk model,
which may be needed to explain the high microlensing optical
depth seen towards the Galactic bulge
\citep{sackett-maxdisk,drim-sperg,bg-disk-mod}.
In a maximum disk model, the rotation curve of the inner part of
the galaxy is almost entirely supported by the stellar mass of the
disk and bulge, so these models predict higher microlensing optical
depths and rates than standard models.
The sixth and seventh lines give microlensing predictions for 
maximum disk models for both the Milky Way and LMC, and the
last line of the table is the total assuming the maximum disk models.

The numbers given in Table~\ref{tab-stars} are identical to those
given in A00, except for the number of expected events, $N_{\rm exp}$,
which has been updated to reflect the correction to the detection
efficiencies that I have presented. Newer models of the LMC
\citep{man-lmc-mod,alves-lmc-mod,gyuk-lmc-mod,nikolaev-lmc-mod,vdMarel-lmc-struc}
give predictions for $\tau$ and $\Gamma$
that are almost identical to the standard model listed in
Table~\ref{tab-stars}.

Comparison of Tables~\ref{tab-taucl} and \ref{tab-stars} reveals that
the measured LMC microlensing optical depth exceeds the prediction
from lensing by known stellar populations of the standard Milky Way and
LMC models at the 99.9\% confidence level, and it exceeds the 
prediction of the maximum disk Milky Way and LMC models at the
99\% confidence level, so it is clear that the basic conclusions of A00
are unchanged when the slight contamination of the candidate microlensing
event sample by variable stars is considered.

A more demanding comparison between the stellar microlensing models
and the data can be achieved by comparing the number of expected events
from each model, $N_{\rm exp}$, with the observed number events. Just 
as in the optical depth comparison, I account for the possibility of variable star 
contamination by using the detected event probability distribution
listed in Table~\ref{tab-true} instead of fixed number of detected events.
An additional factor that must be included is that two of the events in
this sample have additional data and analysis that indicates that these
events are likely to be caused by lenses in the known stellar populations.
These include event 5, where the lens has been identified as an
M-dwarf in the Galactic disk \citep{macho-hstlmc5,dck-lmc5,lmc5-mass},
and event 14, where wiggles in the light curve due to the source star
orbital motion indicate that the lens is most likely to be in the LMC
disk \citep{macho-96lmc2}. (There is also some evidence that the lens
for event 9 may reside in the LMC disk \citep{macho-lmc9}, but this event
did not pass selection criteria A.) These two events are accounted
for with a simple modification to Poisson statistics. The probabilities of
only 1 or 2 events due to lensing by known stellar populations are set
to zero, since we know that there at least 2 such events, and then the
Poisson probabilities for a larger number of events are increased to 
yield a total probability of 1.

The convolution of the detected event  probabilities from Table~\ref{tab-true}
with the modified Poisson statistics on 
$N_{\rm exp}$ for the models then indicates that both the standard
and maximum disk models are about six times less likely to explain
the measured number of events than the measured microlensing
optical depth, $\tau$. Thus, the standard Milky Way and LMC models are
excluded at the 99.8\% confidence level and the maximum disk models
are excluded at the 99.98\% confidence level.

\section{Discussion and Conclusions}
\label{sec-conclude}

In light of recent evidence of contamination of the MACHO LMC
microlensing event sample by variable stars, I've carried out a
new analysis of the MACHO Project 5.7 year LMC data set that
can account for contamination of the microlensing sample due
to imperfect event identification. This results in a slight decrease
in the observed microlensing optical depth to 
$\tau_{\rm meas} = (1.0\pm 0.3) \times 10^{-7}$. If we subtract the
standard model background of $\tau_{\rm back} = 2.4\times 10^{-8}$
and divide by the microlensing optical depth expected for a 
Milky Way dark halo composed entirely of MACHOs 
($\tau_{\rm halo} = 4.7\times 10^{-7}$), we find a most likely
MACHO halo fraction of $f = 0.16 \pm 0.06$, where the error bars
reflect only the microlensing optical depth uncertainty. Additional
uncertainties in halo model parameters will add to this uncertainty,
and of course, it is possible that distribution of the previously
unknown lensing population does not track the distribution of 
the dark halo at all. Nevertheless, the qualitative conclusions
of A00 have not been affected by the modest contamination
of the LMC microlensing sample that has been discovered.
All evidence suggests that the vast majority of the MACHO
LMC microlensing candidates are true microlensing events,
and this result cannot be explained as microlensing by
ordinary stars in known stellar populations.

In estimating the true number of MACHO LMC microlensing events
and the LMC microlensing optical depth, I've made a few conservative
choices. First, I have not corrected for the known systematic 
error that the MACHO selection criteria A discriminates against
non-standard microlensing events such as binary caustic crossing
events like event 9, because the goodness-of-fit to a standard
\pac\ light curve makes up part of the selection criteria. If criteria
A did not discriminate against binary caustic crossing events, then
event 9 might have passed this cut, which would have increased
$\tau$ by $\sim 12$\%.  However, if is not clear that event 9 would
have passed cut A. Also, the frequency of binary caustic crossing
events from the Galactic bulge is only about 5\%, so I expect
that the criteria A bias against such events is likely to be
a systematic error of $< 10$\%. Also, note that this systematic
error implies that the true value of $\tau$ is larger than the 
measured value. \citet{glicens-exotic-tau} discussed a
systematic error for the same type of events with the opposite
sign, but this result does not apply to the event selection criteria
used in A00.

\begin{figure}
\plotone{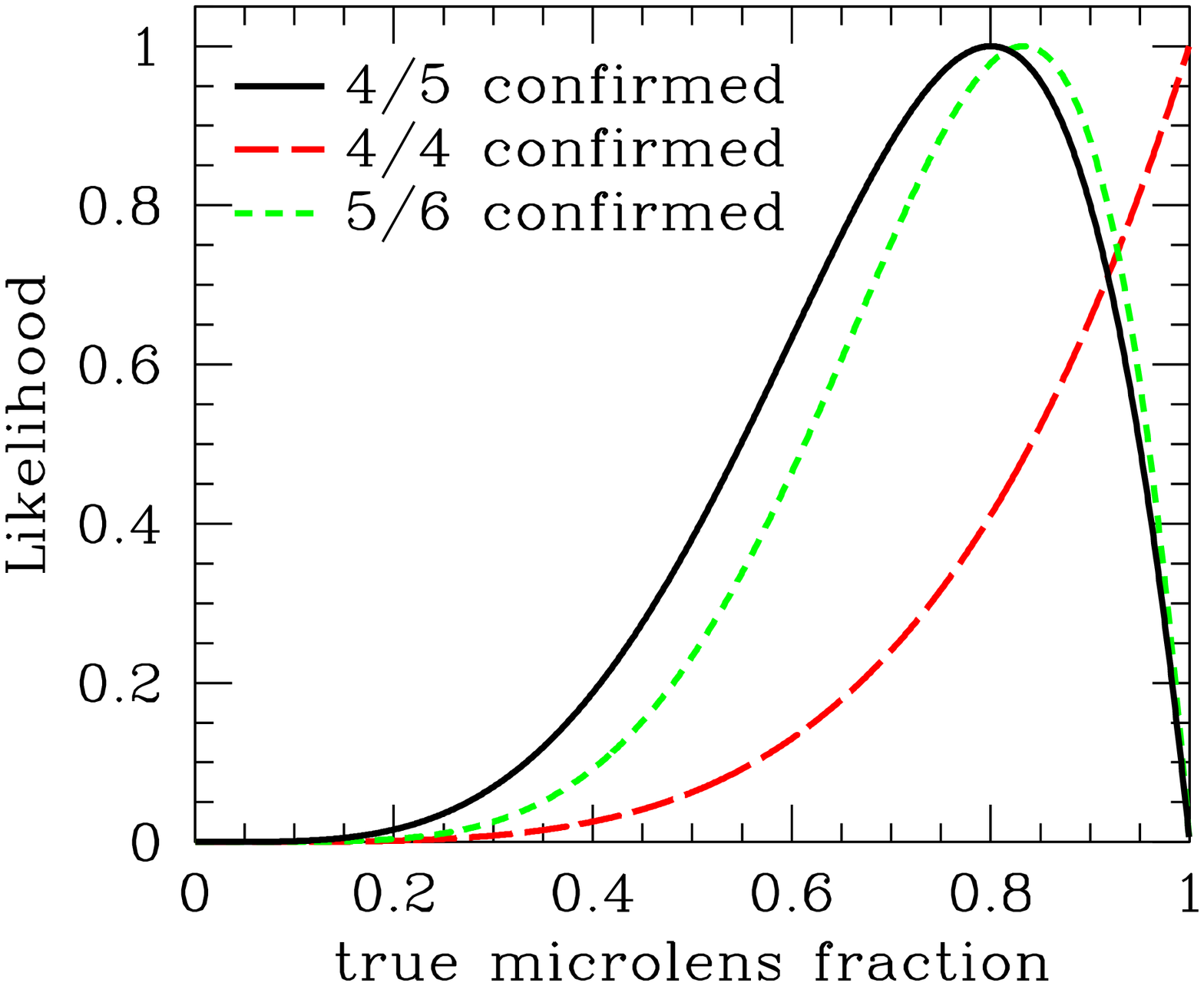}
\caption{
Likelihood functions for the true microlens fraction for the unconfirmed
LMC microlensing event candidates is compared for the different 
microlensing event confirmation scenarios. The 4/5 events confirmed
case is used for the microlensing optical depth and event rate estimates.
\label{fig-mlens_prob}}
\end{figure}

I have also made a conservative choice to include event 23 
as part of the event subset used to estimate the true microlensing
fraction of the unconfirmed events. In fact, it is only the four
events with LMC follow-up observations that represent an
independently selected sub-sample of the full set of events
passing criteria A. Event 23 was added just because it is
the only event which can be rejected from the microlensing
sample. So, by including it, we artificially decrease the
true microlensing fraction plotted in Fig.~\ref{fig-mlens_prob}. This
might be partially compensated by the fact that I've
assigned a microlensing probability of 1 to the confirmed 
events instead of a slightly smaller number.

Thus, it appears that the LMC microlensing puzzle is not
likely to be resolved by the simple experimental error of
the misidentification of variable stars as microlensing events. 
The natural interpretation of the LMC microlensing excess
is that the lens objects comprise part of the Milky Way's dark
halo, which is of unknown composition. Since the measured
microlensing optical depth represents only a fraction ($\sim 16$\%)
of the total halo mass, there is no reason that the distribution of
lens objects must follow that of the bulk of the dark matter. Thus,
the lens objects could follow a distribution like the spheroid
or a very thick disk \citep{gates-gyuk-disk}, and if so, the
total mass of the new population could be significantly less
than 16\% of the dark halo mass.

The timescales of the LMC events suggest a typical lens mass of 
$\sim 0.5\msun$, which suggests that the lenses are likely to
be white dwarfs since main sequence stars of that mass would
be too bright. While white dwarfs were once considered a viable
candidate to comprise the entire dark halo \citep{ros-WDhalo},
there now appear to be significant problems with a previously
unknown population of white dwarfs with enough mass to explain
the LMC microlensing results
\citep{torres-opp-WD,flynn-opp-WD,brook-WDhalosim,MC_halo_WD,diskorhalo_WD},
although some of these constraints are evaded if most of the halo
white dwarfs have Helium atmospheres.

The leading alternative explanation for the LMC microlensing excess
is that the lens objects are ordinary stars associated with the LMC 
\citep{sahu94}, but there is a simple dynamical argument against
this idea \citep{gould-disktau}. On the other hand, the LMC is not an isolated
galaxy and may have had significant dynamical disturbances from the 
Milky Way \citep{weinberg,evans-kerins}, so the assumptions of the
dynamical argument could be wrong. However,
current LMC models that include
these effects still cannot account for the observed microlensing events 
\citep{gyuk-lmc-mod,alves-lmc-mod,man-lmc-mod,nikolaev-lmc-mod}.

A resolution of this puzzle will probably require additional data
so that the distance to a representative sample of LMC lensing
events can be determined. One possibility is to make microlensing
parallax observations from a small telescope in a Heliocentric
orbit \citep{gould-parsat}, such as the 30cm telescope on the
Deep Impact telescope, which could become available for
microlensing parallax observations if an extended mission is
approved. The microlensing
key project for the SIM mission \citep{sim} could also determine
distances to some LMC microlenses.
Microlensing experiments towards
other lines of site, such as M31 could also shed some light on this
issue \citep{m31-mega,m31-pa,m31-halolens,m31-tau}. 
In addition, the SuperMACHO 
\citep{becker-sm}, MOA-II \citep{moa2-hawaii}, and
OGLE-III \citep{ogle3-ews} surveys expect to
substantially increase the detection rate of LMC microlensing
events in order to provide alerts for SIM and DIME and to
measure the spatial variation of the microlensing
optical depth across the face of the LMC.

\acknowledgments

D. P. B. was supported by
grants AST 02-06189 from the NSF and NAG5-13042 from
NASA.

\clearpage


\clearpage

\begin{deluxetable}{rcrrr}  
\tablecaption{Event Classification, Efficiency Corrected $\that$,
and Single Event Optical Depths\label{tab-that} }
\tablewidth{0pt}
\tablehead{
\colhead{Event} &
\colhead{confirmation} &
\colhead{$\that_{\rm nb}$} &
\colhead{$\that_{\rm cor}$} &
\colhead{$\tau_1/10^{-9}$} 
}  
\startdata

  1 & clump giant &   34.2 &  41.9 &  5.4 \qquad \\
  4 & follow-up &   45.4 &  55.5 &  6.3 \qquad  \\
  5 & lens ID &   75.6 &  92.4 &  9.0 \qquad  \\
  6 & unconfirmed &   91.6 & 112.0 & 10.5 \qquad \\
  7 & unconfirmed &  102.9 & 125.8 & 11.6 \qquad \\
  8 & unconfirmed &   66.4 &  81.1 &  8.1 \qquad \\
 13 & follow-up &  100.1 & 122.4 & 11.3 \qquad \\
 14 & follow-up &  100.1 & 122.4 & 11.3  \qquad \\
 15 & follow-up &   36.8 &  45.0 &  5.6 \qquad \\
 18 & unconfirmed &   74.2 &  90.7 &  8.9 \qquad \\
 21 & unconfirmed &   93.2 & 113.9 & 10.7\qquad  \\
 23 & {\it rejected} & - & - & 0 \qquad \\
 25 & clump giant &   85.2 & 104.2 & 9.9 \qquad \\

\enddata
\tablecomments{The quantity $\that_{\rm cor}$ is the average actual event
timescale for events in the A00 Monte Carlo calculations which are detected
with an unblended fit timescale of $\that_{\rm nb}$.  The quantity $\tau_1$ is the
contribution of each event to the total microlensing optical depth,
computed using equation~(\ref{eq-tau1}).  }
\end{deluxetable}

\begin{deluxetable}{ccccc}  
\tablecaption{True Microlensing Event Probabilities\label{tab-true} }
\tablewidth{0pt}
\tablehead{
\colhead{\# of Events} &
\colhead{\bf P(4/5 conf.)} &
\colhead{P(4/4 conf.)} &
\colhead{P(5/6 conf.)} &
\colhead{$\tau_N \times 10^7$}
}  
\startdata
  7 & {\bf 0.013} & 0.004 & 0.008 & 0.59 \\
  8 & {\bf 0.054} & 0.020 & 0.038 & 0.69 \\
  9 & {\bf 0.130} & 0.059 & 0.106 & 0.79 \\
10 & {\bf 0.227} & 0.139 & 0.212 & 0.89 \\
11 & {\bf 0.303} & 0.278 & 0.318 & 0.99 \\
12 & {\bf 0.273} & 0.500 & 0.318 & 1.09 \\
\enddata
\tablecomments{The results of the Likelihood analysis to predict the number
of true microlensing events in the MACHO LMC 5.7 year analysis with 
selection criteria A are compared for different microlensing event
confirmation scenarios. The implied LMC microlensing optical
depth for each number of events is also indicated.
The preferred scenario uses a sub-sample of 5 events
with 4 confirmed as microlensing and one event rejected,
and this is identified in bold face. }
\end{deluxetable}

\begin{deluxetable}{ccccccccccc}
\tablecaption{Optical Depth Confidence Intervals \label{tab-taucl} }
\tablewidth{0pt}
\tablehead{
\colhead{Event Set} & \colhead{\# of} &
  \multicolumn{8}{c}{$\tau (10^{-7})$ \quad for confidence level:} \\
   & \colhead{events} & \colhead{0.001} &
       \colhead{0.01} &  \colhead{0.025} & \colhead{0.05} & \colhead{0.16} &
 \colhead{measured} & \colhead{0.84} & \colhead{0.95} & \colhead{0.975}
}  

\startdata

original-A  & 13 & 0.40 & 0.53 & 0.60 & 0.67 & 0.83 & 1.10 & 1.47 & 1.73 & 1.86\\
{\bf corrected}  & {\bf 7-12} & {\bf 0.24} & {\bf 0.36} & {\bf 0.44} & {\bf 0.50} &
         {\bf 0.66} &  {\bf 0.99} & {\bf 1.32} & {\bf 1.58} & {\bf 1.72}\\
4/4 confirmed & 7-12 & 0.28 & 0.41 & 0.49 & 0.56 & 0.72 & 1.09 & 1.38 & 1.64 & 1.78 \\
\enddata
\tablecomments{ This table compares the microlensing optical
depth $\tau$ in units of $10^{-7}$ for the original MACHO 
selection criteria A analysis with the corrected analysis presented
here.}
\end{deluxetable}

\begin{deluxetable}{lccccc}  
\tablecaption{ Microlensing by Stars
\label{tab-stars} }
\tablewidth{0pt}
\tablehead{
\colhead{Population } &
  \colhead{$\tau (10^{-8}) $} &
  \colhead{$ \VEV{\that}$ (days) } &
  \colhead{$ \VEV{l} (\kpc) $}  &
  \colhead{$ \Gamma (10^{-8} {\rm yr^{-1}}) $}  &
  \colhead{$ \Nexp $} 
}  
\startdata
spheroid                      & 0.19 & 129 & 8.8 & 0.90 & 0.18 \\
thick disk                     & 0.20 & 104 & 3.6 & 0.90 & 0.19 \\[0.10 in]
standard thin disk      & 0.36 & 101 & 1.3 & 1.7 & 0.35  \\
standard LMC disk     & 1.6 & 120 & 50 & 5.8 & 1.21 \\
\tableline
total (min disk)             & 2.35       & - & -    & 9.3 & 1.93 \\
\tableline \\[0.10 in]
maximum thin disk        & 0.59 & 101 & 1.3 & 2.7 & 0.56 \\
maximum LMC disk      & 2.6 & 120 & 50 & 9.8 & 2.05 \\
\tableline
total (max disk)            &  3.58      & - & -     & 14.3 &  2.98 \\

\enddata
\tablecomments{This table shows microlensing quantities for
 various lens populations, with the density and velocity distributions
 and mass functions described in A00.
 $\tau$ is the optical depth, $\VEV{l}$ is the mean lens distance, and
 $\Gamma$ is the total theoretical microlensing rate (see A00).
 The expected number of events $\Nexp$ includes our detection
 efficiency averaged over the $\that$ distribution.
 The LMC values are averaged over the locations of our 30 fields.
 $\Nexp$ is the number of expected events. Two types of disk models
 are considered for the Milky Way: maximum disk models, which have 
 enough mass to account for the rotation curves of each galaxy, and
 standard disk models, which require a massive dark halo made of
 non-microlensing objects to help support the rotation curves.
 }
\end{deluxetable}

\end{document}